\documentclass{ws-ijmpa}

\usepackage{here}

\begin{document}

\markboth{Pawe{\l} Moskal et al.,}
{Hadronic interaction of the $\eta$ meson with two nucleons}

%
\catchline{}{}{}{}{}
%

\title{HADRONIC INTERACTION OF THE $\eta$ MESON WITH TWO NUCLEONS}

\author{\footnotesize P.~MOSKAL$^{\ast,\dagger}$, H.-H.~ADAM$^\ddagger$,
                      A.~BUDZANOWSKI$^{\dagger \dagger}$, 
                      R.~CZY{\.Z}YKIEWICZ$^{\ast,\dagger}$, D.~GRZONKA$^\dagger$, M.~JANUSZ$^\ast$, 
                      L.~JARCZYK$^\ast$, B.~KAMYS$^\ast$, A.~KHOUKAZ$^\ddagger$,
                      K.~KILIAN$^\dagger$, P.~KLAJA$^\ast$, J.~MAJEWSKI$^{\ast,\dagger}$,
                      W.~OELERT$^\dagger$, C.~PISKOR--IGNATOWICZ$^\ast$,
                      J.~PRZERWA$^\ast$,
                      T.~RO{\.Z}EK$^{\star,\dagger}$, T.~SEFZICK$^\dagger$,
                      M.~SIEMASZKO$^\star$, J.~SMYRSKI$^\ast$,
                      A.~T{\"A}SCHNER$^\ddagger$, P.~WINTER$^\dagger$, M.~WOLKE$^\dagger$, 
                      P.~W{\"U}STNER$^{\ddagger \ddagger}$, W.~ZIPPER$^\star$ }
\address{
$^\ast$ Nuclear Physics Department, Jagellonian University, Cracow, 30-059, Poland\\
$^\dagger$ Institut f{\"u}r Kernphysik, Forschungszentrum J{\"u}lich, J{\"u}lich, 52425, Germany\\
$^\ddagger$ Institut f{\"u}r Kernphysik, Universit{\"a}t M{\"u}nster,M{\"u}nster, 48149, Germany\\
$^{\dagger \dagger}$ Institute of Nuclear Physics, Cracow, 31-342, Poland\\
$^\star$ Institute of Physics, University of Silesia, Katowice, 40-007, Poland\\
$^{\ddagger \ddagger}$ ZEL Forschungszentrum J{\"u}lich, J{\"u}lich, 52425, Germany\\
}

\maketitle

\pub{Received (Day Month Year)}{Revised (Day Month Year)}

\begin{abstract}

The COSY-11 collaboration has conducted experiments aiming at the determination
of the excitation function and phase-space population of the $pp\to pp\eta$ reaction
close to the kinematical threshold. The precise data obtained with the stochastically
cooled proton beam of the cooler synchrotron COSY and the high resolution zero-degree
magnetic spectrometer allowed for the observation of the significant deviations 
--~in the shape 
of the excitation function and two-particle invariant masses~-- 
from the 
predictions based on the assumption that the reaction phase space is homogenously populated.
Comparison of the shape of the excitation function for the $pp\to pp\eta$ and $pp\to pp\eta^{\prime}$
reaction allows to distinquish in the model independent way an influence originating 
from the proton-proton and proton-$\eta$ interaction. For the comparison the full data set from 
experiments performed at COSY and other laboratories is used.


\end{abstract}

\vspace{0.3cm}

\section{Introduction}
In analogy to the connection between electromagnetic and Van der Vaals potentials,
we may perceive the hadronic force
as a residuum of the strong interaction that occurs between quarks and gluons -- the
constituents of hadrons.
Therefore, the knowledge of the interaction of hadrons is interesting not only on its own account 
but also since it delivers information about the
structure of hadrons and the strong interaction itself.

The fact that fourty years after the discovery of the $\eta$ and $\eta^{\prime}$ 
mesons~\cite{pevsner421}
their interactions
with nucleons remain so weakly established,
indicates that it is rather challenging to conduct research which could deliver information about
this interaction.
The scattering length
--~the very basic quantity describing the low energy interaction potential~-- in the case of the $\eta$ meson
is poorly estimated,
and in the case of the $\eta^{\prime}$ meson it is entirely unknown.
Estimated value of the real part of the proton-$\eta$ scattering length varies from 0.25~fm to 1.05~fm
depending on the approach employed
for its 
determination~\cite{greenwycech04}.

The main obstacle in the experimental studies 
involving any 
of {\em neutral} ground state mesons 
is too short life--time 
which prohibits their utilization as secondary beams. Therefore the study of
their interaction with hadrons is accessible only via observations of their
influence on the cross section of the reactions in which they were produced
(eg. $NN \rightarrow NN\, Meson$).
The influence of the relatively weak nucleon-$\eta$ interaction may be magnified when
producing the meson in the vicinity of two nucleons. In this context the $pp\eta$ system reveals to be particularly
interesting since neither the $pp$ nor the $p\eta$ interaction is strong enough to form a bound state and so the
$pp\eta$ system may occur to be Borromean.
It was pointed out by Wycech~\cite{wycech2981} that the large enhancement in the excitation function 
of the $pp\to pp\eta$ reaction observed close 
to the kinematical threshold can be described assuming that 
the proton-proton pair is produced from a large object of a 4~fm radius.
Yet, at present  
it is still not established 
whether the low energy $pp\eta$ system can really form a Borromean  or resonant state.
Though the significant progress in the understanding
of the production mechanism on the hadronic 
level
has been achieved~\cite{wilkinjohansson,vadim024002,kanzoheber,bernard259},
the final state system was always treated approximately utterly ignoring the 
$\eta$-proton final state interaction.
Only, very recently the rigorous three-body approach  has been applied for the  description 
of the observed  excitation function and two-particle invariant mass spectra of the $pp\to pp\eta$ 
reaction. Two independent approaches by Deloff~\cite{deloffc11} 
and Fix \& Arenh{\"o}vel~\cite{fixnew}
revealed that the rigorous three body treatment of the final state leads to the results significantly
different from the  two-body approach even when the $\eta$-proton interaction is completely
disregarded. Also a very important conclusion drawn in reference~\cite{fixnew} is that the role of
the final state interaction depends crucially on the range of the primary reaction dynamics,
showing that the inaccuracy of establishing  contributions from the exchange of various mesons 
limits the inferences about the interaction among the final state particles.

The data which stimulated the development
of the abovementioned  three body formalism 
will be presented hereafter.
For the better visualisation of the effects due to the proton-$\eta$ interaction we will compare
the shape of the excitation functions of the $pp\to pp\eta$  and $pp\to pp\eta^{\prime}$ reactions
and will discuss the observed differences.
In the next section which 
constitutes an extraction from the more comprehensive treatise~\cite{hab}
we will present only the excitation function and for the details concerning the differential 
distributions the reader is referred to the more 
exhaustive 
experimental~\cite{hab,moskal025203,TOFeta}
and  theoretical elaborations~\cite{vadim024002,fixnew,nakayama0302061,deloff}.

\section{Excitation function of the reactions $pp\to pp\eta$ and $pp\to pp\eta^{\prime}$}
\begin{figure}[H]
    \parbox{0.4\textwidth}{\centerline{
    \includegraphics[width=0.4\textwidth]{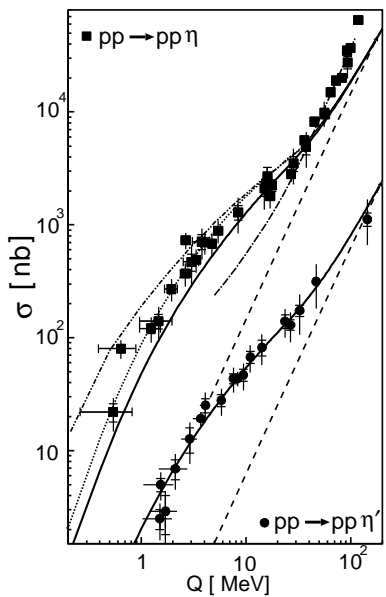}}}
    \hfill
    \parbox{0.56\textwidth}{
    \vspace{-0.5cm}
    \caption{\label{cross_eta_etap}
      Total cross section for the reactions
      $pp \rightarrow pp \eta^{\prime}$~(circles) and
      $pp \rightarrow pp \eta$~(squares)
      as a function of the centre-of-mass
      excess energy Q. Data are from
      refs.~\protect\cite{moskal025203,balestra29,hibou41,bergdoltR2969}
      The dashed lines
      indicate a phase space integral normalized arbitrarily. 
      The solid lines
      show the phase space distribution with inclusion of the
      $^1S_0$ proton-proton strong
      and Coulomb interactions.
    In case of the $pp\to pp\eta$ reaction the solid line
    was fitted to the data in the excess energy range
    between 15 and $40\,\mbox{MeV}$. Additional
    inclusion of the proton-$\eta$ interaction is indicated by the dotted line.
    The scattering length of $a_{p\eta} = 0.7\,\mbox{fm} + i\,0.4\,\mbox{fm}$ and the
    effective range parameter 
    $b_{p\eta} = -1.50\,\mbox{fm} -i\,0.24\,\mbox{fm}$~\protect\cite{greenR2167} 
    have been chosen arbitrarily.
    The dashed-dotted line represents the energy dependence taking into account the
    contribution from the
    $^3P_{0}\to ^1\!\!\!S_{0}s$, $^1S_{0}\to ^3\!\!\!P_{0}s$ and
    $^1D_2\to ^3\!\!\!P_2 s$ transitions~\protect\cite{nakayama0302061}.
    Preliminary results for the $^3P_{0}\to ^1\!\!\!S_{0}s$
    transition with full treatment of three-body effects
    are shown as a dashed-double-dotted line~\protect\cite{fixnew}.
    The absolute scale of dashed-double-dotted line
    was arbitrary fitted to demonstrate
    the energy dependence only.
    }}
  \end{figure}

The determined energy dependences of the total cross section
  for $\eta^{\prime}$~\cite{balestra29,hibou41} 
  and
  $\eta$~\cite{moskal025203,hibou41,bergdoltR2969} 
  mesons production in  proton-proton collisions
  are presented in figure~\ref{cross_eta_etap}.
  Comparing the data to the arbitrarily normalized phase space integrals (dashed lines)
  reveals that the $pp\eta$ FSI enhances the total cross section by more than an order
  of magnitude for low excess energies.
\vspace{-0.2cm}
\vspace{-0.3cm}
  One recognizes also that
  in the case of the $\eta^{\prime}$ meson
  the data are described very well (solid line)
    assuming that the on-shell proton-proton amplitude
    exclusively determines the phase space population.
In the case of $\eta$ meson production the interaction between nucleons is
evidently not sufficient to describe the increase of the total cross section
for very low and very high excess energies, as can be concluded from the comparison
of the data and the upper solid line in figure~\ref{cross_eta_etap}.
This line was normalized to the data at an excess energy
range between $15\,\mbox{MeV}$ and $40\,\mbox{MeV}$.
The enhancement of the total cross section for higher energies can be assigned to
the outset of higher partial waves,
and the  discrepancy visible closer to the threshold can be plausibly explained by the influence of the
attractive interaction between the $\eta$ meson and the proton.
 A similar effect close-to-threshold is also observed
  in the data of photoproduction of $\eta$
  via  the $\gamma d\to pn\eta$ reaction~\cite{eta_photo} indicating
  to some extent that the phenomenon is independent of the production
  process 
 and rather is related to the interaction among the $\eta$ meson and
  nucleons. 

  The dotted--line in figure~\ref{cross_eta_etap} corresponds to the 
simple phenomenological treatment~\cite{review}
   based on the factorization of the
  transition amplitude into the constant primary production and the on-shell
  incoherent pairwise interaction among outgoing particles.
  Although it describes
  the enhancement close-to-threshold very well,
  it fails to describe the invariant
  mass distribution of the proton-proton and proton-$\eta$ subsystems
  determined recently at Q~=~15~MeV by the COSY-TOF~\cite{TOFeta}
  and at Q~=~15.5~MeV by the COSY-11~\cite{moskal025203} collaborations.
  It was suggested in reference~\cite{nakayama0302061} that the bump in the
  invariant mass spectra may be due to the contribution from higher partial waves.
  However, the amount of the P-wave admixture 
  derived from the proton-proton invariant mass
  distribution  
  spoils significantly
  the agreement with  the data at low values of Q
  (see dashed-dotted line in figure~\ref{cross_eta_etap}).
  Another explanation proposed is based on the anticipation that the
  production amplitude may vary with energy~\cite{deloff}.
 However, both listed approaches neglected the
  proton-$\eta$ interaction and only now the first calculations
  with the rigorous treatment of the three-body final state  including 
  both proton-proton and proton-$\eta$ interaction are available.
  And although biased with neglection of the Coulomb effect~\cite{fixnew}
  or imaginary part of the proton-$\eta$ scattering length~\cite{deloffc11} 
  they herald a fully rigorous
   explanation of the observed spectra. 

\vspace{-0.3cm}

\section*{Acknowledgments}
 The work has been supported 
 by the DAAD Exchange Programme (PPP-Polen) and
 by the Polish State Committee for Scientific Research
 (grant No. PB1060/P03/2004/26).

\end{document}